\documentclass[twocolumn,prl,aps,showpacs]{revtex4}
\usepackage{epsfig}
\usepackage[english]{babel}
\usepackage{latexsym}
\usepackage{graphics}
\usepackage{subfigure}

\begin{document}

\title{Dynamics of a Bright Soliton in Bose-Einstein condensates with
Time-Dependent Atomic Scattering Length in an Expulsive Parabolic Potential}
\author{Z. X. Liang$^{1,2}$, Z. D. Zhang$^{1}$, W. M. Liu$^{2}$}

\address{$^1$Shenyang National Laboratory for Materials Science, Institute of Metal
Research, and International Centre for Materials Physics, Chinese
Academy of Sciences,\\ Wenhua Road 72, Shenyang 110016, People's Republic of China\\
$^2$ Institute of Physics, Chinese Academy of Sciences, Beijing
100080, People's Republic of China}

\begin{abstract}
We present a family of exact solutions of one-dimensional
nonlinear Schr\"odinger equation, which describe the dynamics of a
bright soliton in Bose-Einstein condensates with the
time-dependent interatomic interaction in an expulsive parabolic
potential. Our results show that, under the safe range of
parameters, the bright soliton can be compressed into very high
local matter densities by increasing the absolute value of atomic
scattering length, which can provide an experimental tool for
investigating the range of validity of the one-dimensional
Gross-Pitaevskii equation. We also find that the number of atoms
in the bright soliton keeps dynamic stability: a time-periodic
atomic exchange is formed between the bright soliton and the
background.
\end{abstract}

\maketitle

With the experimental observation and theoretical studies of
Bose-Einstein condensates (BECs)\cite{Dalfovo}, intensive
interests have been paid to the nonlinear excitations of the
atomic matter waves, such as dark\cite{Burger} and bright
solitons\cite{Strecker,Khaykovich,Kevreidis,Perez-Garcia}. It is
believed that the atomic matter bright solitons are of primary
importance for developing concrete applications of BEC in the
future, so it is of interest to develop a new technique which
allows us construct a particular soliton with the assumed peak
matter density. One possibility is to vary the interatomic
interaction by means of external magnetic fields. Recent
experiments have demonstrated that the variation of the
ef\/fective scattering length, even including its sign, can be
achieved by utilizing the so-called Feshbach resonance \cite
{Roberts}. This of\/fers a good opportunity for manipulation of
atomic matter waves and nonlinear excitations in BEC
\cite{Pelinovsky,Liang}. In Ref. \cite{Abdullaev}, it has been
demonstrated that the variation of nonlinearity of the
Gross-Pitaevskii (GP) equation via Feshbach resonance provides a
powerful tool for controlling the generation of bright and dark
soliton trains starting from periodic waves. Besides, a sinusoidal
variation of the scattering length has also been used to form
patterns such as Faraday waves \cite{Staliunas}, or as a means to
maintain BECs without an external trap in two dimensions
\cite{Saito}.

In this letter, we present a thorough analysis on the dynamics of
a bright soliton of BEC with time-varying atomic scattering length
in an expulsive parabolic potential. Our study is greatly
facilitated by the so-called Darboux transformation
\cite{Matveev}, by which we can directly construct the exact
solutions of one-dimensional(1D) nonlinear Schr\"odinger
equation(NLSE). Under the safe range of parameters, the bright
soliton in BEC can be compressed into very high local matter
densities by increasing the absolute value of atomic scattering
length with Feshbach resonance. During the compression of the
bright soliton in BEC, the number of atoms in the bright soliton
keeps dynamic stability and the exchange of the atoms between the
bright soliton and the background of density exists.

At the mean-field level, the GP equation governs the evolution of
the macroscopic wave function of a 3D BEC. In the physically
important case of the cigar-shaped BEC, it is reasonable to reduce
the GP equation into 1D NLSE\cite{Perez-Garcia, Kevrekidis,
Brazhnyi},
\begin{eqnarray}  \label{GP}
i\frac{\partial \psi\left(x,t\right)}{\partial t}+\frac{\partial
^{2}\psi\left(x,t\right)}{\partial
x^{2}}&+&2a(t)|\psi\left(x,t\right)|^{2}
\psi\left(x,t\right)  \nonumber \\
&+&\frac{1}{4}\lambda ^{2}x^{2}\psi\left(x,t\right)=0.
\end{eqnarray}
In Eq. (\ref{GP}), time $t$ and coordinate $x$ are measured in
units $ 2/\omega_{\bot}$ and $a_{\bot}$, where
$a_{\bot}=(\hbar/m\omega_{\bot})^{1/2} $ and
$a_{0}=(\hbar/m\omega_{0})^{1/2}$ are linear oscillator lengths in
the transverse and cigar-axis directions, respectively.
$\omega_{\bot}$ and $ \omega_{0}$ are corresponding harmonic
oscillator frequencies, $m$ is the atomic mass and
$\lambda=2|\omega_{0}|/\omega_{\bot}\ll1$. The Feshbach-managed
nonlinear coefficient reads $a\left( t\right)=|a_{s}(t)|/a_{B}
=g_{0}\exp \left( \lambda t\right)$($a_{B}$ is the Bohr
radius)\cite{P¨¦rez-Garc¨ªa, Sc}. The normalized macroscopic wave
function $\psi\left(x,t\right)$ is connected to the original order
parameter $\Psi(\mathbf{r},t)$ as follows,
\begin{eqnarray}  \label{order parameter}
\Psi\left(\mathbf{r},t\right)&=&\frac{1}{\sqrt{2\pi a_{B}}a_{\perp
}} \psi\left(\frac{x}{a_{\perp }},\frac{\omega _{\perp
}t}{2}\right)  \nonumber\\
&&\times\exp\left(-i\omega_{\perp
}t-\frac{y^{2}+z^{2}}{2a_{\perp}}\right).
\end{eqnarray}

From the viewpoint of stability, 3D and 1D equations are very
different. For a true 1D system, one does not expect the collapse
of the system with increasing number of atoms\cite{Perez-Garcia}.
However, it happens that a realistic 1D limit is not a true 1D
system, with the density of particles still increasing due to the
strong restoring forces in the perpendicular directions. To avoid
the collapse of the bright soliton\cite{Fibich}, we must restrict
our study of BEC to the safe range of parameters, in which the
system becomes effectively 1D, i. e. the energy of two body
interactions is much less than the kinetic energy in the
transverse direction\cite{Brazhnyi}:
$\epsilon^{2}=a_{\bot}/\xi^{2}\sim N |a_{s}|/a_{0}\ll 1$ ($\xi$ is
the healing length). The bright soliton in BECs has been created
with the parameters of $N\approx\times10^{3}$ , $\omega _{\perp
}=2\pi \times 700Hz$ and $\omega _{0}=2\pi \times 7Hz$,
$a_{final}=-4a_{B}$ for $^{7}Li$ \cite{Khaykovich}, which provides
the safe range of parameters. With the same experimental
conditions in Ref. \cite{Khaykovich} and $a_{s}(t=0)=-0.25a_{B}$,
we can calculate $\epsilon^{2}=a_{\bot}/\xi^{2}\sim N |a_{s}|/
a_{0}=9.5\times10^{-3}\ll 1$. Then, the scattering length is
increased in the form of $a\left( t\right)=g_{0}\exp \left(
\lambda t\right)$. After at least up to $50$ dimensionless units
of time, the absolute value of the atomic scattering length turns
to $|a_{s}(t)| =0.8a_{B} $, corresponding to
$\epsilon^{2}=a_{\bot}/\xi^{2}\sim N
|a_{s}|/a_{0}=3\times10^{-2}\ll 1$. Under the above conditions,
the system is effectively 1D. So the safe range of parameters can
be described as follows: (1) with the same experimental conditions
in Ref. \cite{Khaykovich}; (2) ramp up the scattering length in
the form of $a\left( t\right)=g_{0}\exp \left( \lambda t\right)$
within $50$ dimensionless units of time. We also have to specify
the terminology long-time dynamics. A unity of time, $
\bigtriangleup t=1$, in the dimensionless variables corresponds to
$2/\omega _{\perp }$ real seconds. This means, for example, that
for a BEC in a trap with transversal size of the order of
$a_{\perp }\approx 1.5\mu m$, a unity of the dimensionless time
corresponds to $5.0\times10^{-4}s$. The lifetime of a BEC in
today's experiments is of the order of $1s $, which  is about
$200$ in our dimensionless units.

The so-called `seed' solution of Eq. (\ref{GP}) can be chosen as
follows,
\begin{equation}  \label{Ac}
\psi _{0}\left( x,t\right) =A_{c} \exp \left[ \frac{\lambda t}{2}+
i\varphi _{c}\right] ,
\end{equation}
where $\varphi _{c}=k_{0}x\exp (\lambda t) -\frac{\lambda
x^{2}}{4} +\frac{( 2g_{0}A_{c}^{2}-k_{0}^{2}) ( \exp( 2\lambda t )
-1) }{2\lambda }$ and $A_{c}$ and $k_{0}$ are the arbitrary real
constants. We perform the Darboux transformation\cite{Darboux} $
\psi _{1}=$ $\psi_{0} +\frac{2}{\sqrt{g_{0}}} \frac{\left( \zeta
+\overline{\zeta }\right) \phi _{1}\overline{\phi _{2}}}{ \phi
^{T}\overline{\phi }}\exp \left( -\lambda t/2-i\lambda
x^{2}/4\right) $ to obtain the new solution of Eq. (\ref{GP}) by
taking Eq. (\ref{Ac}) as the seed. Then we obtain the exact
solution of Eq. (\ref{GP})\ as follows:
\begin{eqnarray}  \label{solution}
\psi &=&\left[A_{c}\!+\!A_{s}\frac{(\gamma \cosh \theta \!+\!\cos
\varphi )\!+\!i(\alpha \sinh \theta \!+\!\beta \sin \varphi
)}{\cosh \theta +\gamma
\cos \varphi }\right]  \nonumber \\
&&\times \exp \left(\frac{\lambda t}{2}+i\varphi_{c}\right),
\end{eqnarray}
where
\begin{eqnarray}
\theta &=&-\frac{\left[ \left( k_{0}+k_{s}\right) \Delta
_{R}-\sqrt{g_{0}} A_{s}\Delta _{I}\right] \left[ \exp \left(
2\lambda t\right)-1\right] }{
2\lambda }  \nonumber \\
&&+\Delta _{R}x\exp \left( \lambda t\right) ,  \nonumber \\
\varphi &=&-\frac{\left[ \left( k_{0}+k_{s}\right) \Delta
_{I}+\sqrt{g_{0}} A_{s}\Delta _{R}\right]\left[ \exp \left(
2\lambda t\right) -1\right] }{
2\lambda }  \nonumber \\
&&+\Delta _{I}x\exp \left( \lambda t\right),
\end{eqnarray}
and
\begin{eqnarray}
\alpha &=&\frac{\sqrt{g_{0}}A_{c}\left( k_{0}-k_{s}+\Delta
_{I}\right) }{\Lambda },  \nonumber \\
\beta &=&1-\frac{2g_{0}A_{c}^{2}}{\Lambda },  \nonumber \\
\gamma &=&\frac{\sqrt{g_{0}}A_{c}\left( \Delta _{R}-\sqrt{g_{0}}
A_{s}\right) }{\Lambda },
\end{eqnarray}
with
\begin{eqnarray}  \label{MI}
\Delta &=&\sqrt{\left[ -\sqrt{g_{0}}A_{s}+i\left(
k_{s}-k_{0}\right) \right]
^{2}-4g_{0}A_{c}^{2}}\equiv \Delta _{R}+i\Delta _{I},  \nonumber \\
\Lambda &=&g_{0}A_{c}^{2}+\frac{\left( \Delta _{R}-\sqrt{g_{0}}
A_{s}^{2}\right) }{4}+\frac{\left( k_{s}-k_{0}+\Delta _{I}\right)
^{2}}{4},
\end{eqnarray}
where $k_{s}$ is the arbitrary real constant. On the one hand,
when $A_{c}=k_{0}=0$, Eq. (\ref{solution}) reduces to the
well-known soliton solution: $\psi _{s}=A_{s}\text{sech}\theta
_{s}\exp (\lambda t/2+i\varphi _{s})$, where $\theta
_{s}=-\sqrt{g_{0}}\exp (\lambda
t)A_{s}x+\sqrt{g_{0}}k_{s}A_{s}\left[ \exp \left( 2\lambda
t\right) -1\right] /\lambda $ and $\varphi _{s}=\varphi _{c}-
g_{0}A_{c}^{2}\left[\exp\left(2\lambda t\right) -1\right]
/2\lambda $. On the other hand, when the amplitude of this soliton
vanishes ($A_{s}=0$) , Eq. (\ref{solution}) reduces to Eq.
(\ref{Ac}). Thus, Eq. (\ref{solution}) represents a bright soliton
embedded in the background. Considering the dynamics of the bright
soliton on the background, the length $2L$ of the spatial
background must be very large compared to the scale of the
soliton. In the real experiment \cite{Strecker}, the length of
background of BEC can be reached at least $2L=$370$\mu m$. At the
same time, in Fig. 1, the width of the bright soliton is about
$2l=2\times1.4$ $\mu m$=2.8 $\mu m$( a unity of coordinate,
$\Delta x=1 $ in the dimensionless variables, corresponds to
$a_{\bot}=(\hbar/m\omega_{\bot})^{1/2} =1.4\mu m$). So, we indeed
have $l\ll L$, a necessary condition for realizing our soliton in
experiment.

By utilizing the properties of Eq. (\ref{solution}), we
demonstrate that the manipulation of the scattering length can be
used to compre ss a bright soiton of BEC into an assumed peak
matter density. It has been reported that an abrupt change of the
scattering length can lead to the splitting of the soliton with
generating the new solitons. The fragmentation of the soltion
obviously decreases the numbers of atoms of the original soliton,
which is undesirable for application\cite{Carr}. However, in Eq.
(\ref{solution}), the change of the scattering length preserves
the soliton from splitting new ones. For simplicity, we assume
$k_{0}=k_{s}$ in Eq. (\ref{MI}) and only consider the case of
$A_{s}^{2}>4A_{c}^{2}$, for in the case of $A_{s}^{2}<4A_{c}^{2}$
, a small perturbation for Eq. (\ref{solution}) may lead to the
modulation instability \cite{Salasnich2}. On the conditions above,
Eq. (\ref{solution}) can be deduced into the following form:

\begin{eqnarray}  \label{norm}
\psi &=&\left[-A_{c}+\delta_{2}\frac{\delta _{2}\cos \varphi
-iA_{s}\sin
\varphi }{A_{s}\cosh \theta -2A_{c}\cos \varphi }\right]  \nonumber \\
&&\times \exp \left(\frac{\lambda t}{2}+i\varphi _{c}\right),
\end{eqnarray}
where
\begin{eqnarray}
\theta &=&\sqrt{g_{0}}\delta _{2}x\exp (\lambda
t)-\frac{\sqrt{g_{0}}
k_{0}\delta _{2}\left[ \exp (2\lambda t)-1\right]}{\lambda },  \nonumber \\
\varphi &=&-\frac{g_{0}A_{s}\delta _{2}\left[ \exp(2\lambda
t)-1\right]}{2\lambda } ,  \nonumber \\
\delta _{2} &=&\sqrt{A_{s}^{2}-4A_{c}^{2}}.
\end{eqnarray}
\begin{figure}[htbp]
\begin{center}
\epsfxsize=8cm\epsffile{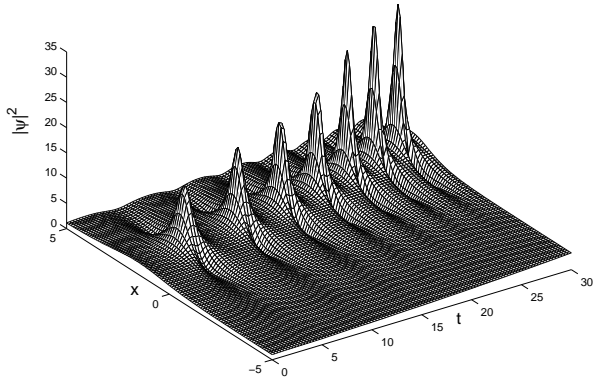}
\end{center}
\caption{The dynamics of Feshbach resonance managed soliton in the
expulsive parabolic potential given by Eq. (\protect\ref{norm}).
The parameters are given as follows:$
\protect\lambda=2\times10^{-2}$,
$g_{0}=\frac{\protect\sqrt{2}}{2}$, $A_{c}=1 $, $A_{s}=2.4$,
$k_{0}=0.03$, }
\end{figure}
For a better understanding, we plot Eq. (\ref{norm}) in Fig. 1,
which shows the dynamics of the Feshbach resonance managed bright
soliton in the expulsive parabolic potential. As one can see from
Fig. 1, with the increasing absolute value of the scattering
length, the bright soliton has an increase in the peaking value,
and a compression in the width. As a result, we can get a bright
soliton with the assumed peak matter density. It is interesting to
observe that in the expulsive parabolic potential, the bright
soliton is set into motion and propagates in the longitudinal
direction, instead of oscillating in an attractive parabolic
potential. The possibility of compressing BEC solitons into an
assumed peak matter density could provide an experimental tool for
investigating the range of validity of the 1D GP equation. Since
the quasi-1D GP equation only applies for low densities, it would
indeed be interesting to see how far one can compress a soliton in
a real experiment by increasing the a bsolute value of the
scattering length.

Inspired by two experiments\cite{Strecker, Khaykovich}, we can
design the compression of a bright soliton in BECs with the
following steps: (i) create a bright soliton in BECs with the
parameters of $N\approx\times10^{3}$ , $\omega _{\perp }=2\pi
\times 700Hz$ and $\omega _{0}=2\pi \times 7Hz$, $
a_{s}=-0.25a_{B}$ for $^{7}Li$. The main effect of this expulsive
term is that the center of the BEC accelerates along the
longitudinal direction.  (ii)under the safe range of parameters
discussed above, ramp up the absolute value of the scattering
length according to $a(t)=g_{0}exp(\lambda t)$ due to Feshbach
resonance, where $\lambda=2|\omega_{0}|/\omega_{\bot}=2
\times10^{-2}$ is a very small value. A unity of time,
$\bigtriangleup t=1$ , in the dimensionless variables corresponds
to $2/\omega _{\perp }=4.5\times10^{-4}$ real seconds. (iii) after
at least up to $50$ dimensionless units of time, the absolute
value of the atomic scattering length turns to $|a_{s}(t)|
=0.8a_{B} $, which is less than $|a_{final}|=4a_{B}$. This means
that during the process of compressing the bright soliton, the
stability of soliton and the validity of 1D approximation can be
kept as displayed in Fig. 1. Therefore, the phenomena discussed in
this letter should be observable within the current experimental
capability.

Furthermore, based on Eq. (\ref{norm}), we find that when $\sinh
\theta =0$ , the peak matter density of bright soliton arrives at
the maximum
\begin{equation}
\left\vert \psi \right\vert ^{2}=\exp \left( \lambda t\right)
\left( A_{c}^{2}+\frac{\delta _{2}A_{s}}{A_{s}-2A_{c}\cos \varphi
} \right)
\end{equation}
and when $\cosh \theta =\frac{A_{s}}{A_{c}\cos \varphi }-
\frac{A_{c}\cos \varphi }{A_{s}},$ the peak matter density of
bright soliton arrives at the minimum
\begin{equation}
\left\vert \psi \right\vert ^{2}=\exp \left( \lambda t\right)
\left( A_{c}^{2}-\frac{A_{c}^{2}\delta _{2}^{2}\cos ^{2}\varphi }{
A_{s}^{2}-4A_{c}^{2}\cos ^{2}\varphi }\right)
\end{equation}
This means that the bright soliton can only be squeezed into the
assumed peak matter density between the minimum and maximum
values. In order to investigate the stability of the bright
soliton against the variation of the scattering length in the
expulsive parabolic potential, we obtain
\begin{equation}  \label{Num}
\lim_{L\rightarrow \infty }\label{Num} \int_{-L }^{+L }\left[
\left\vert \psi \left( x,t\right) \right\vert ^{2}-\left\vert \psi
\left( \pm L ,t\right) \right\vert ^{2} \right] dx=\frac{2\delta
_{2}}{\sqrt{g_{0}}}.\nonumber
\end{equation}
which is the exact number of the atoms in the bright soliton
against the background described by Eq. (\ref{norm}). This
indicates that during the process of the compression of the bright
soliton, the number of atoms in the bright soliton keeps
invariant. In contrast, the quantity
\begin{eqnarray}  \label{exnum}
\kappa&=&\lim_{L\rightarrow \infty }\int_{-L }^{+L }\left\vert
\psi \left( x,t\right) -\psi
\left(\pm L ,t\right) \right\vert ^{2}dx  \nonumber \\
&=&\frac{2\delta _{2}}{\sqrt{g_{0}}}\left( 1+A_{c}M\cos
\varphi\right),
\end{eqnarray}
where
\begin{equation}
M=\frac{4\arctan \frac{\sqrt{A_{s}+2A_{c}\cos \varphi }}{\sqrt{
A_{s}-2A_{c}\cos \varphi }}}{\sqrt{A_{s}^{2}-4A_{c}^{2}\cos ^{2}\varphi }}
\end{equation}
counts number of atoms in both the bright soliton and background
under the condition of $\psi(\pm L ,t)\!\neq\! 0$. Eq.
(\ref{exnum}) displays that a time-periodic atomic exchange is
formed between the bright soliton and the background. As shown in
Fig. 2(a), in the case of zero-background, i. e. $A_{c}=0 $, there
will not be the exchange of atoms, However, in the case of
nonzero-background as shown in the Figs. 2 (b) and (c), the
exchange of atoms between the bright soliton and the background
becomes more quickly with increasing the absolute value of the
scattering length. The conclusion can be made that the number of
atoms in the bright soliton in BEC keeps dynamic stability against
the variation of the scattering length in the expulsive parabolic
potential.
\begin{figure}[htbp]
\begin{center}
\epsfxsize=8cm\epsffile{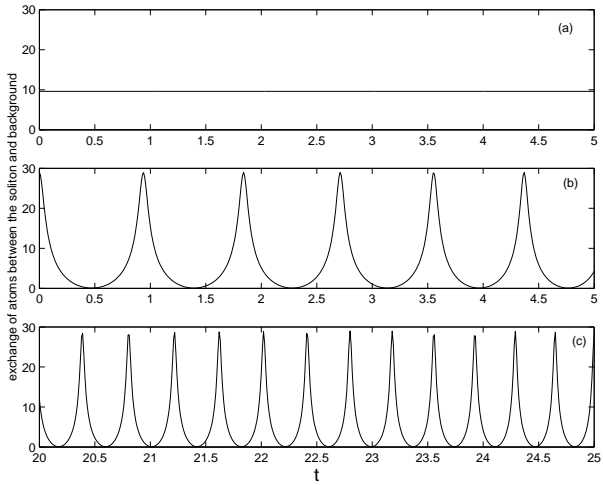}
\end{center}
\caption{Time-periodic atomic exchange between the bright soliton
and the background given by Eq. (\protect\ref{exnum}). The range
of time is: (a) and (b) t=[0, 5], (c) t=[20, 25]. The parameters
are given as follows: $\protect\lambda=0.02$, $g_{0}=1$,
$A_{s}=4.8$, (a) $A_{c}=0$ , (b) and (c) $A_{c}=2.3$.}
\end{figure}
The consideration provided above implies that we should construct
the bright soliton in BEC on the background of density described
by Eq. (\ref{Ac}). A question arises about the possibility of
creation of such a state experimentally. We notice that Eq.
(\ref{norm}) will take the particular form at the time
$t_{0}=\frac{1}{2 \lambda}\ln \left[ -\frac{\lambda\left(
4n+1\right) \pi }{A_{s}\delta _{2}g_{0}}+1\right]
,n=-1,-2,-3...$as follows
\begin{eqnarray}  \label{combin}
\psi\left(x,t\right) =&-&A_{c}\exp \left(\frac{\lambda t_{0}}{2}
\right)\exp\left(i\varphi _{c}\right)  \nonumber \\
&&-i\delta _{2}\exp \left(\frac{\lambda t_{0}}{2}\right)\text{sech}\theta
\exp \left(i\varphi _{c}\right),  \nonumber \\
=&-&\psi _{0}\left( x,t_{0}\right) -\psi _{soliton}\left( x,t_{0}\right) .
\end{eqnarray}
which is the linear combination of Eq. (\ref{Ac}) and a bright
soliton. So Eq. (\ref{combin}) means that Eq. (\ref{norm}) can be
generated by coherently adding a bright soliton into the
background of density described by Eq. (\ref{Ac}).

In conclusion, we present a family of exact solutions of the
nonlinear Schr\"odinger equation with the time-varying nonlinear
coefficient in the expulsive parabolic potential. Our results
describe the dynamics of Feshbach resonance managed bright soliton
of BEC in an expulsive parabolic potential. Furthermore, under the
safe range of parameters, it is possible to squeeze a bright
soliton of BEC into the assumed peak matter density, which can
provide an experimental tool for investigating the range of
validity of the 1D GP equation. We also find that the number of
atoms in the bright soliton keeps dynamic stability: the exchange
of atoms between the bright soliton and the background becomes
more quickly with increasing the absolute value of the scattering
length. Recent developments of controlling the scattering length
in the experiments allow for the experimental investigation of our
prediction in the future.

We would like to thank the referees for numerous useful
suggestions. We also express our sincerely thanks to Biao Wu and
Lan Yin for helpful discussions. This work is supported by the NSF
of China under Grant Nos.50331030, 10274087, 90103024 and
10174095.


\begin{thebibliography}{99}
\bibitem{Sc} Because of $\lambda\ll 1$, the nonlinear coefficient
$a(t)=g_{0}e^{\lambda t}$ can be expressed as
$a(t)=g_{0}(1+\lambda t+0^2(\lambda t))$.
\bibitem{Darboux} Considering the following UV-pair $\phi_{x}=U \phi$
, $\phi_{t}=V \phi$,  $\phi=( \phi_{1,}\phi_{2}) ^{T}$ where
$U=\zeta J+P$, $V=2i\zeta^{2}J+ \lambda x \zeta J+2i\zeta P+W$,
with $J=\left(\matrix{1 & 0 \cr 0 & -1 }\right), P=\left(
\matrix{0 & \sqrt{g_{0}}Q \cr -\sqrt{g_{0}}\overline{Q} & 0
}\right)$, $W=\left( \matrix{ig_{0}|Q|^{2} & -\sqrt{g_{0}}\lambda
x Q+i\sqrt{g_{0}}Q_{x} \cr \sqrt{g_{0}}\lambda x
\overline{Q}+i\sqrt{g_{0}}\overline{Q}_{x} & -ig_{0}|Q|^{2}
}\right)$. Here the overbar denotes the complex conjugate and
$\psi=Q \exp\left(-i\lambda x^{2}/4-\lambda t/2\right )$. From the
compatibility condition $U_{t}-V_{x}+\left[U, V\right]=0$ one can
derive Eq. (\ref{GP}) in the case of $\zeta=\xi\exp{(\lambda t)}$.
\end{thebibliography}
\end{document}